\documentclass[aps,prb,english,twocolumn,superscriptaddress]{revtex4-1}
\usepackage{graphicx}
\usepackage{amsfonts}
\usepackage{amsmath}
\usepackage{amssymb}
\usepackage{babel}

\begin{document}

\title{First-principles study of crystal and electronic structure of rare-earth cobaltites}

\author{M. Topsakal}
\affiliation{Department of Chemical Engineering and Materials Science, University of Minnesota, Minneapolis, Minnesota 55455, USA}
\author{C. Leighton}
\affiliation{Department of Chemical Engineering and Materials Science, University of Minnesota, Minneapolis, Minnesota 55455, USA}
\author{R.~M. Wentzcovitch}
\affiliation{Department of Chemical Engineering and Materials Science, University of Minnesota, Minneapolis, Minnesota 55455, USA}

\date{\today}

\begin{abstract}

Using density functional theory plus self-consistent Hubbard $U$ (DFT$+U_{sc}$) calculations, we have investigated the structural and electronic properties of the rare-earth cobaltites \textit{R}CoO$_3$ (\textit{R} = Pr -- Lu). Our calculations show the evolution of crystal and electronic structure of the insulating low-spin (LS) \textit{R}CoO$_3$ with increasing rare-earth atomic number (decreasing ionic radius), including the invariance of the Co-O bond distance ($d_{Co-O}$), the decrease of the Co-O-Co bond angle ($\Theta$), and the increase of the crystal field splitting ($\Delta_{CF}$) and  band gap energy ($E_g$). Agreement with experiment for the latter improves considerably with the use of DFT$+U_{sc}$ and all trends are in good agreement with experimental data. These trends enable a direct test of prior rationalizations of the trend in spin-gap associated with the spin crossover in this series, which is found to expose significant issues with simple band based arguments. We also examine the effect of placing the rare-earth \textit{f}-electrons in the core region of the pseudopotential. The effect on lattice parameters and band structure is found to be small, but distinct for the special case of \textit{Pr}CoO$_3$ where some \textit{f}-states populate the middle of the gap, consistent with recent reports of unique behavior in Pr-containing cobaltites. Overall, this study establishes a foundation for future predictive studies of thermally induced spin excitations in rare-earth cobaltites and similar systems.

\end{abstract}

\maketitle

\section{Introduction} \label{sec:intro}

Rare-earth cobaltites with general formula \textit{R}CoO$_3$ (\textit{R} = rare-earth) form an intriguing family of strongly correlated perovskites that have been the subject of extensive research.\cite{1953,1967,goodenough58,demazeau74,rmp} They exhibit unique changes in electronic, magnetic, and thermal properties as a function of temperature (T),\cite{T-1,T-2,T-3,T-4} strain ($\epsilon$),\cite{e-1,e-2,han-prb85} ion substitution on the \textit{R}-site (x),\cite{X-1,X-2,X-3} etc. It has been challenging to understand these properties from strictly experimental or theoretical standpoints, as many of these property changes are closely related to a subtle phenomenon, the spin crossover, i.e. the total electron spin ($S$) of Co$^{3+}$ changing with the above-mentioned factors. For applications, \textit{La}CoO$_3$ has been recognized as promising contact material for the cathode in solid oxide fuel cells (SOFCs).\cite{sofc1,sofc2} It has also been reported that Sr substituted \textit{La}CoO$_3$ can act as a fast ion conducting material and as a high temperature oxygen separation membrane.\cite{membrane1,membrane2} 

Among the \textit{R}CoO$_3$ compounds, lanthanum cobaltite (\textit{La}CoO$_3$) has been a model system for investigation of  the spin crossover, whose nature has been debated since the late 50's.\cite{goodenough58} There are six 3\textit{d} electrons in the Co$^{3+}$ ions of \textit{R}CoO$_3$ perovskite and hence the possible electron spin states are S=0, 1, 2, conventionally referred to as low-spin (LS; $t_{2g}^{6}e_{g}^{0}$), intermediate-spin (IS; $t_{2g}^{5}e_{g}^{1}$) and high-spin (HS; $t_{2g}^{4}e_{g}^{2}$), respectively. This is atomic-like language, however, which is not a fully appropriate description,\cite{lee2013} a point that will be returned to below.

At low temperatures, the \textit{R}CoO$_3$'s are insulating and Co$^{3+}$ ions are in the low-spin (LS,S=0) ground state. Depending on the crystal-field splitting ($\Delta_{CF}$) between the $e_g$-$t_{2g}$ states and the Hund exchange energy ($\Delta_{EX}$), LS Co$^{3+}$ ions can be excited to either "IS" or "HS" states, i.e. the spin crossover can occur. Thermal expansion,\cite{lco_thm1} magnetic susceptibility,\cite{lco_sus1} and heat capacity\cite{lco_heat1} measurements mark an onset spin-state transition temperature ($T_{onset}$) to a paramagnetic insulating phase at around 30 K for \textit{La}CoO$_3$. As the temperature is further increased to $\sim$ 530 K, a second transition to a metallic phase is observed by distinct anomalies in thermal expansion,\cite{lco_thm2} magnetic susceptibility,\cite{lco_sus1} heat capacity,\cite{lco_heat2} and resistivity\cite{lco_res} measurements. The spin-state in the intermediate temperature range is still highly debated.\cite{LSIS1,LSIS2,LSHS1,LSHS2,han-prb82,han-prb85,demkov,leighton2015} 

Of most relevance to this work, magnetic susceptibility measurements for \textit{(Pr,Nd)}CoO$_3$,\cite{lco_sus1} \textit{Eu}CoO$_3$,\cite{eucoo3} and \textit{Lu}CoO$_3$\cite{lucoo3} reveal an interesting variation of $T_{onset}$, which increases to $\sim$197, $\sim$299, $\sim$400 and $\sim$535 K for \textit{Pr}CoO$_3$, \textit{Nd}CoO$_3$, \textit{Eu}CoO$_3$, and \textit{Lu}CoO$_3$, respectively. This variation is commonly attributed to a change of Co-O-Co bond angles and $\sigma^*$-bonding $e_g$ bandwidth.\cite{series,goodenough69,goodenough71} Specifically, due to the observed independence of the Co-O bond distance with \textit{R}-ion, $\Delta_{CF}$ is typically assumed constant. The increase in $T_{onset}$ from La-Lu is thus viewed as being due to an increase in the spin gap due to the $e_{g}$ bandwidth decreasing, the $t_{2g}$ bandwidth being ignored. While this provides simple semi-quantitative rationalization of this trend,\cite{series} suitably detailed electronic structure calculations could provide a more rigorous test, which seems to not yet have been performed.  We believe that a comprehensive first-principles study accross the \textit{R}CoO$_3$ series, where structural and electronic properties are well documented, is essential for guiding further studies, thus clarifying the true nature of this evolution. 

To this end, here we report : i) a consistent investigation of basic structural and electronic properties throughout the entire rare-earth cobaltite series, with particular emphasis placed on “trends” along the series; ii) a survey of performance of different exchange correlation functionals in calculating these properties; iii) the influence of rare-earth 4\textit{f} electrons on these properties. Such investigations are fundamental to establish a foundation for future studies in this intriguing class of perovskite oxides, particularly on the phenomenon of spin crossovers along this series. With regard to the latter, we expose significant limitations with prior simple rationalizations in the trend of $T_{onset}$  across this series, suggesting that much more detailed and rigorous calculations are required. Our work forms a basis for such.

\section{Methods} \label{sec:method}

All calculations were performed using the Quantum ESPRESSO software.\cite{qe} Structural optimizations for 20-atom \textit{R}CoO$_3$ cells with $Pbnm$ symmetry, at various pressures, were performed with variable cell shape molecular dynamics,\cite{VC-relax} and a $6 \times 6 \times 4$ shifted \textbf{k}-point mesh with a Fermi-Dirac smearing factor of 0.002 Ry. Equation of state curves were fitted to the third-order Birch-Murnaghan equation.\cite{B-M} To improve the visibility, density of states (DOS) plots were smoothed using a Gaussian with 0.003 Ry width. For Co and O ions, we use the projector augmented-wave (PAW) data sets;\cite{pslib} a 50 (400) Ry wave function (charge density) cutoff was used. For rare-earth elements, we generated special ultra-soft pseudopotentials.\cite{vb-uspp} In the all-electron computation part of the pseudopotential generation scheme, the \textit{f}-shell filling was chosen as appropriate for the trivalent state: two for Pr, three for Nd, etc., through to fourteen for Lu. Then the 4\textit{f} electrons were considered as core states for the pseudopotential generation part. The consequences of placing \textit{R}CoO$_3$ 4\textit{f} electrons into the core are further investigated for selected \textit{R}CoO$_3$'s in Sec III (C) below.

For strongly correlated 3\textit{d} electrons on the Co site, the on-site Coulomb interaction can be treated using the DFT+$U$ functional:\cite{matteo2005, burakprb, dudarev}
\begin{equation}
E_{DFT+U}=E_{DFT}[\rho(\textbf{r})]+\sum\limits_{I,\sigma}\frac{U^I}{2}\text{Tr}[\textbf{n}^{I\sigma}(1-\textbf{n}^{I\sigma})],
\label{eq:DFT+U}
\end{equation}
where $E_{DFT}[\rho(\textbf{r})]$ is the DFT energy as a functional of the electron density $\rho(\textbf{r})$, $U^I$ is the Hubbard $U$ parameter to treat the on-site Coulomb interaction of atomic site $I$, and $\textbf{n}^{I,\sigma}$ is 
the occupation matrix of the atomic site $I$ with spin index $\sigma$. In this paper, we compute the Hubbard $U$ in Eq.~(\ref{eq:DFT+U}) of the LS Co using a self-consistent procedure;\cite{burakprb, Campo10, Hsu-prl, kulik} the resultant Hubbard $U$ is referred to as the self-consistent $U$ ($U_{sc}$) hereafter. A detailed description of this procedure can be found in Ref.~\onlinecite{Hsu-prl} and its online supplemental material. In brief, we start with a DFT+$U$ calculation with a trial $U$ ($U_{in}$) to obtain the desired spin states. By applying local perturbations to the Co site in the DFT+$U_{in}$ ground state, with the Hubbard potential being held fixed, the second derivative of the DFT energy with respect to the electron occupation at the Co site can be obtained using linear response theory.\cite{matteo2005} This second derivative, $U_{out}$, will be used as $U_{in}$ in the next iteration. Such a procedure is repeated until self-consistency is achieved, namely, $U_{in}=U_{out} \equiv U_{sc}$. It should be noted that the Hubbard $U$ parameters of the \textit{R}CoO$_3$'s reported here were obtained using the specific approach \cite{matteo2005, kulik, Hsu-prl, Campo10} implemented in the Quantum ESPRESSO package. Other DFT+U codes with different implementations will likely give  somewhat different results using these U-values. Nevertheless, the present calculation and use of \textit{U} are consistent. This approach has been successfully applied to numerous problems, including in calculations of spin crossovers in minerals at high pressures and temperatures.\cite{Tsuchiya-prl2006,Hsu-epsl,Hsu-2014-prb}

\section{Results and Discussion} \label{sec:results}

\subsection{Structural properties of \textit{R}CoO$_3$} \label{subsec:struct}

\begin{figure*}
\includegraphics[width=10cm]{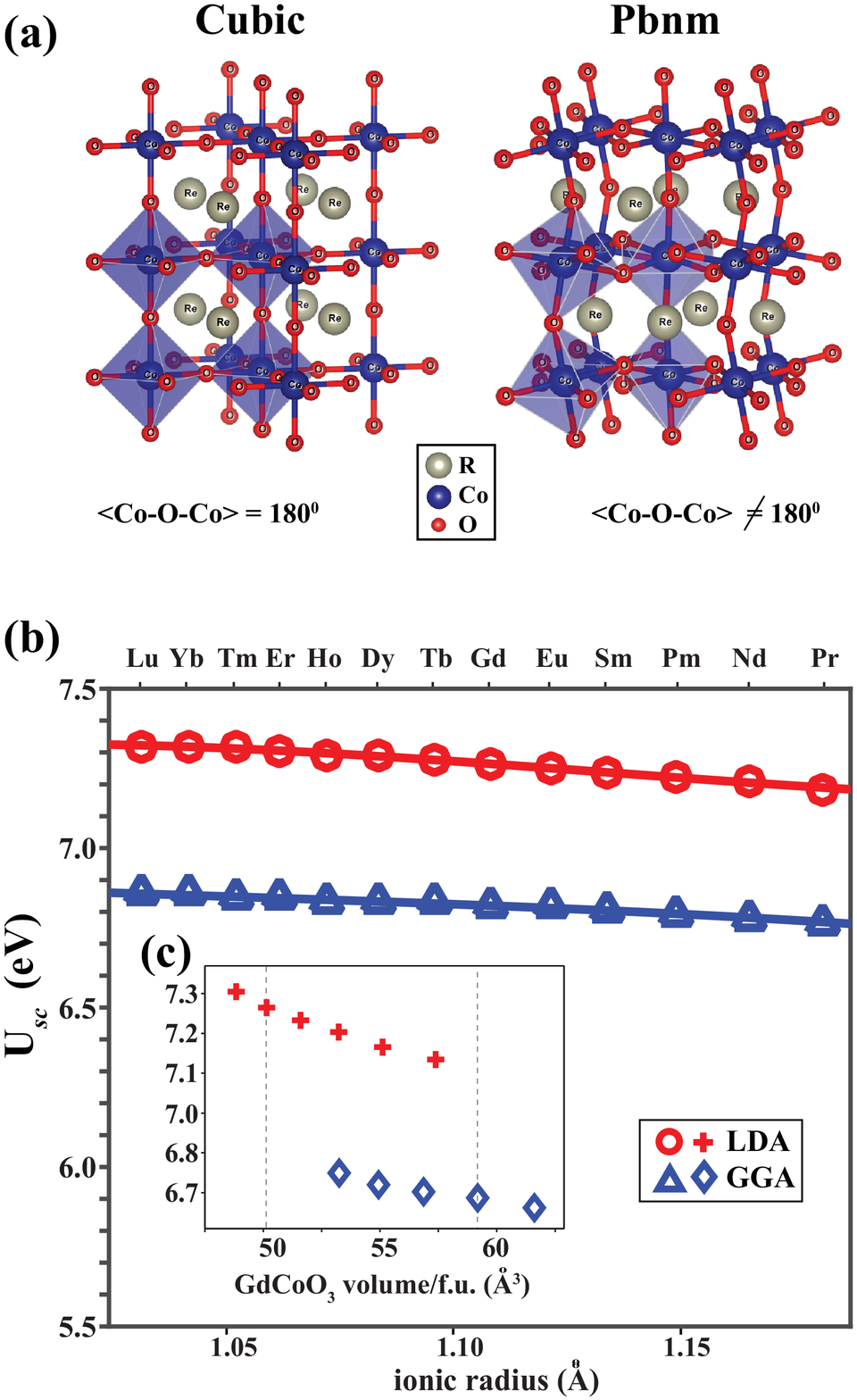}
\caption{(Color online) (a) Cubic and \textit{Pbnm} perovskite structures of \textit{R}CoO$_3$. Larger dark (blue) and light (grey) spheres are cobalt and rare-earth ions respectively. Oxygen ions are denoted by smaller red spheres; (b) calculated self-consistent Hubbard $U$ values of low-spin (LS) Co in \textit{R}CoO$_3$ along the series from Lu to Pr. Red (circle) and blue (triangle) symbols were obtained with LDA and GGA functionals and the continuous line is a guide for the eye; (c) variation of $U_{sc}$ with volume for \textit{Gd}Co$_3$. Dashed vertical lines show zero-pressure equilibrium volumes calculated with LDA and GGA functionals.}
\label{fig:Us}
\end{figure*}

All \textit{R}CoO$_3$ perovskites have a Goldschmidt tolerance factor\cite{goldschmidt} smaller than one. Their crystal structures are distorted perovskites, no longer in cubic symmetry. Among them, \textit{La}CoO$_3$ has rhombohedral ($R\overline{3}c$) symmetry, while the others have $Pbnm$ symmetry [($a^-a^-c^+$) in Glazer notation], forming a 20-atom unit cell, as illustrated in Fig.~\ref{fig:Us}(a). As a result of octahedral rotation, the Co-O-Co bond angles are no longer 180$^0$ in the $Pbnm$ structure. Because \textit{La}CoO$_3$ has different crystal symmetry, and has previously been investigated extensively, we omit \textit{La}CoO$_3$ in this study.  

To examine the performance of widely used exchange-correlation (XC) functionals, we computed the structural parameters of \textit{R}CoO$_3$ using the local density approximation (LDA), generalized gradient approximation (GGA), LDA+$U_{sc}$, and GGA+$U_{sc}$. The self-consistent Hubbard $U$ of low-spin Co ($U_{sc}$) for both LDA and GGA are shown in Fig.~\ref{fig:Us}(b). As can be seen, $U_{sc}$ slightly increases with the atomic number, i.e. $U_{sc}$ increases with decreasing ionic radius or unit cell volume. Such a trend is consistent with the slight increase of $U_{sc}$ with hydrostatic pressure observed in iron-bearing minerals.\cite{Hsu-epsl,Hsu-2014-prb} Fig.~\ref{fig:Us}(c) shows the variation of $U_{sc}$ at different volumes for \textit{Gd}CoO$_{3}$, as an example. The variation of $U_{sc}$ with volume is relatively small, again similarly to iron-bearing minerals.\cite{Hsu-epsl,Hsu-2014-prb}

\begin{figure*}
\includegraphics[width=16.5cm]{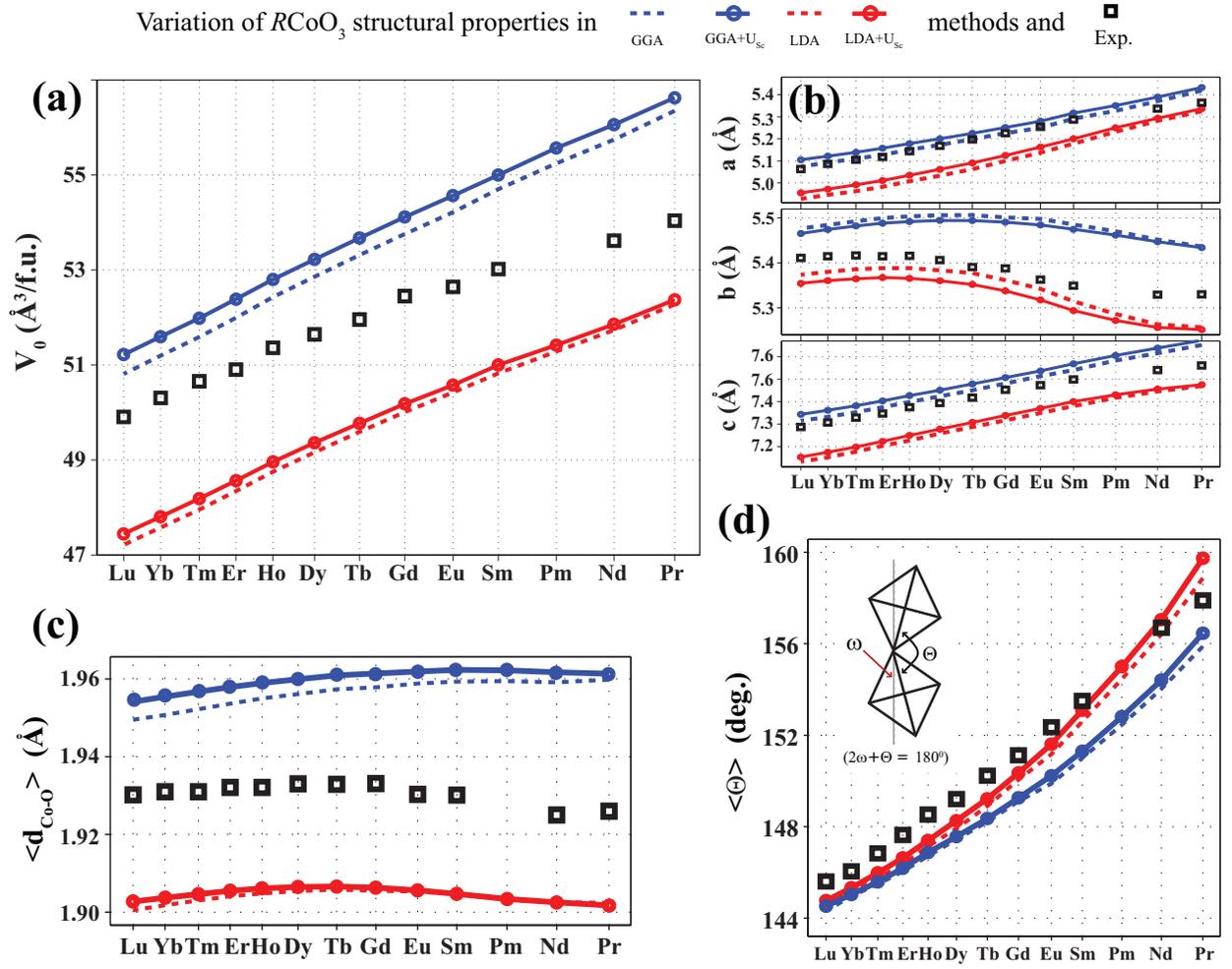}
\caption{(Color online)  Structural parameters of \textit{R}CoO$_3$'s. Experimental values from Ref. \onlinecite{series} are indicated by empty squares while calculated values with LDA/GGA and LDA+U$_{sc}$/GGA+U$_{sc}$ methods are presented as dashed lines and circles respectively. Red color is used for LDA and LDA+U$_{sc}$ while blue color is used for GGA and GGA+U$_{sc}$ calculations. (a), Equilibrium volume, V$_0$ , per formula unit; (b) lattice parameters of the 20-atom orthorhombic Pbnm perovskite structure at zero  pressure; (c) average Co-O bond lengths; (d) Co-O-Co bond angles.}
\label{fig:struct}
\end{figure*}

In Fig.~\ref{fig:struct} we show the computed structural parameters along with the experimental data, including the equilibrium volume ($V_0$), the $Pbnm$ lattice parameters ($a$, $b$, $c$), the average Co-O distance ($\langle d_{Co-O} \rangle$), and the average Co-O-Co bond angle ($\langle \Theta \rangle$). It is known that LDA usually underestimates lattice constants, while the GGA functional usually overestimates them, providing lower and upper bounds for the predicted volume. As shown in Fig.~\ref{fig:struct}(a), the equilibrium volumes observed in experiments are indeed within the range bounded by LDA and GGA calculations. For both cases, the computed $V_0$ increases with the inclusion of the Hubbard $U$. The $Pbnm$ lattice parameters ($a$, $b$, $c$) are presented in Fig.~\ref{fig:struct}(b). Noticeably, as the cation size of \textit{R} and unit-cell volume increase, the lattice parameters $a$ and $c$ increase steadily, while $b$ first increases slightly then decreases. These trends are clear in both experimental and calculated cases. Fig.~\ref{fig:struct}(c) shows, however, that the resulting average Co-O distance $\langle d_{Co-O} \rangle$ is nearly constant along the rare-earth series. Instead, the effect of the \textit{R} cation radius is manifested in terms of increasing octahedral rotation (away from 180$^o$) with decreasing \textit{R} cation radius (see Fig.~\ref{fig:struct}(d)). 

Throughout the series \textit{R}=Pr -- Lu, all the considered functionals thus successfully predict the evolution of structural parameters. Among them, our calculations suggest that the LDA+$U_{sc}$ is the most appropriate for treating \textit{R}CoO$_3$. While LDA+$U_{sc}$ yields equilibrium volumes smaller than experiment, this underestimation can be further improved by a few percent with the inclusion of vibrational effects.\cite{w-pnas,RiMG-rw1} In contrast, the overestimation of volume in GGA and GGA+$U_{sc}$ cannot be improved; inclusion of zero-point motion would only make the prediction less accurate. We thus favor LDA+$U_{sc}$ for these materials.

\subsection{Electronic structure of \textit{R}CoO$_3$} \label{subsec:electronic}

\begin{figure*}
\begin{center}
\includegraphics[width=17cm]{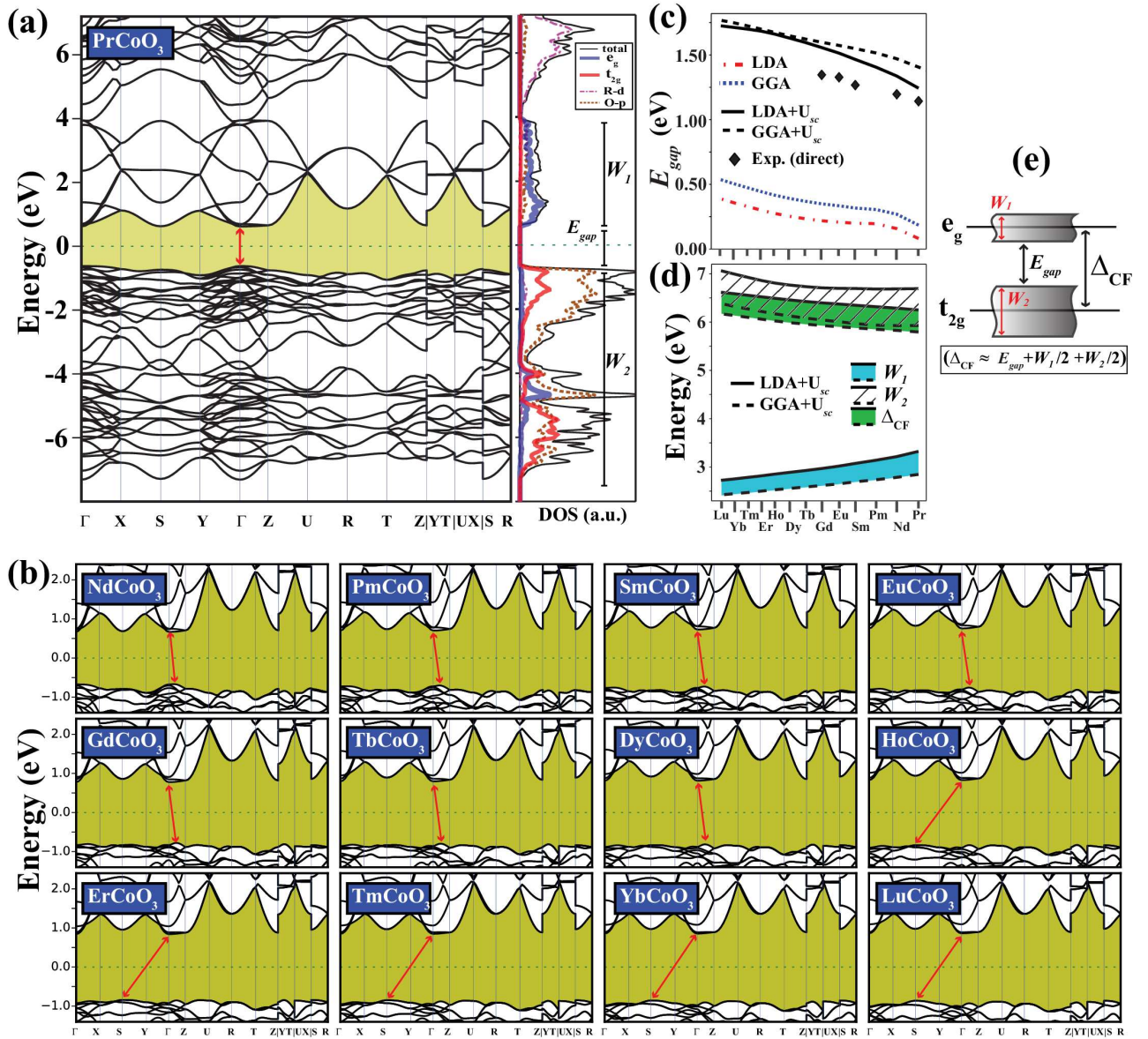}
\caption{(Color online) (a) LDA+U$_{sc}$ calculated band structure and projected density of states of \textit{Pr}CoO$_3$ along high-symmetry lines of the orthorhombic Pbnm lattice. Zero of the energy was set at the Fermi level. The band gap is shaded and red arrows connect the maximum in these valence bands and the maximum in these conduction bands; (b) LDA+U$_{sc}$ band structures of other \textit{R}CoO$_3$'s; (c) variation of band gaps (\textit{E$_{gap}$}) along the series calculated using different methods with some experimental direct band gaps taken from Ref. \onlinecite{yamaguchi}; (d) variation of $\sigma^*$-bonding $e_g$ bands (\textit{$W_1$}) and occupied O \textit{p} and $t_{2g}$ bands (\textit{$W_2$}),  and crystal field splitting energy ($\Delta_{CF}$) between $e_{g}$-$t_{2g}$ orbitals. Continuous lines show LDA+U$_{sc}$ values and dashed lines show GGA+U$_{sc}$ values with the area between them shaded in color or texture;  (e) diagram relating these quantities in \textit{R}CoO$_3$'s. }
\label{fig:dos}
\end{center}
\end{figure*}

The band structure of \textit{Pr}CoO$_3$ (as an example) determined using the favored LDA+U$_{sc}$ is plotted along the high-symmetry lines in Fig.~\ref{fig:dos}(a). A direct band gap of 1.24 eV at the $\Gamma$ point can be observed. The states at the top of the valence band have mainly O 2\textit{p} character with some Co $t_{2g}$ contribution, as can be seen in the density of states (DOS) plot on the right side of Fig.~\ref{fig:dos}(a). However, the states at the bottom of the conduction band have mainly Co e$_g$ character. In ideal octahedral environment, all six 3\textit{d} electrons should occupy $t_{2g}$ orbitals. As a result of octahedral rotations and of hybridization between O \textit{p} and cobalt e$_g$ states, there is some e$_g$ contribution to valence states. While the states in the conduction band ranging from 0.5 to 4 eV have mainly e$_g$ character, there is also some O \textit{p} contribution. The combination of these empty O \textit{p} and Co \textit{d} states are referred to as $\sigma^*$-bonding $e_g$ bands.\cite{series}  The rare-earth 5\textit{d} states of \textit{Pr}Co$_3$ lie 3.5 eV above the conduction band minimum and do not contribute to the bands near the Fermi level. The calculated band structures of the other \textit{R}CoO$_3$'s are presented in Fig.~\ref{fig:dos}(b) for the energy window of -1.5 to 2.5 eV from the Fermi level. In contrast to \textit{Pr}CoO$_3$, the other \textit{R}CoO$_3$'s have indirect band gaps. The conduction band minimum is at $\Gamma$ for all \textit{R}CoO$_3$. From \textit{Pr}CoO$_3$ to \textit{Dy}CoO$_3$, the valence band maximum occurs along the high symmetry line connecting $\Gamma$ and $Z$ points. Although the difference is within a few meV, beneath the expected accuracy of methods used here, the valence band maximum shifts to the $S$ point for the rest of \textit{R}CoO$_3$'s.

Key changes in the electronic structure of \textit{R}CoO$_3$ for \textit{R}=Pr -- Lu are captured in Figures ~\ref{fig:dos}(c,d,e). The band gap (\textit{E$_{gap}$}) increases by $\sim0.48$ eV ($\sim39$\%) along the series from Pr to Lu. For some \textit{R}CoO$_3$'s, experimental\cite{yamaguchi} direct band gap values are shown in Fig.~\ref{fig:dos}(c). Although our calculated band gap values are indirect, importantly, the increasing trend of band gaps from \textit{Pr}CoO$_3$ to \textit{Gd}CoO$_3$ is well reproduced by our calculations. It should also be emphasized in this respect that, as shown in Fig.~\ref{fig:dos}(c) for \textit{Pr}CoO$_3$, \textit{Nd}CoO$_3$, \textit{Sm}CoO$_3$, \textit{Eu}CoO$_3$ and \textit{Gd}CoO$_3$, LDA-only and GGA-only calculations severely underestimate the band gaps compared to experimental values. Therefore, the electronic properties are not properly described in calculations without U, although the effect of adding \textit{U} is small with regard to structural parameters.

Moving on from the band gap, Fig. ~\ref{fig:dos}(d) further shows that the $\sigma^*$ ($e_g$-derived) bandwidth ($W_1$), the $t_{2g}$ bandwidth ($W_2$), and the crystal field splitting ($\Delta_{CF}$), are also dependent on the rare-earth cation. The definitions of these parameters are illustrated in Figures ~\ref{fig:dos}(a,e) and their relation with $E_{gap}$ is clarified in Fig. ~\ref{fig:dos}(e). It should be noted here that we define $\Delta_{CF}$ as the splitting between the midpoints of the $e_g$- and $t_{2g}$-derived bands, and that in Fig. ~\ref{fig:dos}(d) both the LDA+U$_{sc}$ and GGA+U$_{sc}$ values are shown for each quantity, with the region between them shaded or colored. Examining the results, from Pr through Lu, $W_1$ decreases (by $\sim$0.5 eV), while $W_2$ and $\Delta_{CF}$ increase, by $\sim$0.45 and $\sim$0.37 eV, respectively. It is revealing to compare these values and trends to simple expectations. While $W_1$ indeed decreases with increasing deviation from 180$^o$ Co-O-Co bond angles (Fig. ~\ref{fig:struct}(d)), it should be noted that $\Delta_{CF}$ is \textit{not} constant as might naively be expected from the approximately constant Co-O bond length (Fig. ~\ref{fig:struct}(c)). In addition, $W_2$ , which might naively be viewed as negligible in comparison to $W_1$  in a simple picture,\cite{goodenough71} actually significantly exceeds $W_1$. This latter feature has been seen in several first principles electronic structure calculations.\cite{width-other1, width-other2} 

 It is insightful to consider the impact of the above on simple arguments made for the evolution of $T_{onset}$, as summarized in the Introduction. Within a simple treatment, this onset temperature for the spin crossover can be viewed as proportional to a spin gap energy,\cite{goodenough71} $E_{sg} \approx \Delta_{CF} - W_2/2 -W_1/2 - \Delta_{EX}$, where $\Delta_{EX}$ is the intra-atomic Hund exchange energy of Co in an octahedral environment. As mentioned in the Introduction, in the picture advanced by Tachibana \textit{et. al.},\cite{series} the increasing spin gap with decreasing rare-earth ionic size is then rationalized by assuming an invariant $\Delta_{EX}$ and $\Delta_{CF}$, a negligible $W_2$, and a $W_1$ that decreases from Pr to Lu. Though we have not calculated $\Delta_{EX}$, the clear issues from our calculated values are that $W_2$ is not negligible (it in fact exceeds $W_1$ and also depends on the rare-earth cation), and that $\Delta_{CF}$, contrary to simple expectations, is not constant. It thus appears that the simple rationalization of the trend in $T_{onset}$­ across the series \textit{Pr}CoO$_3$ to \textit{Lu}CoO$_3$ is not straightforwardly borne out by a detailed electronic band structure picture. However, simply inserting band structure results for these quantities into the simple approximated expression for the spin gap given above is likely hazardous. Specifically, this approach does not capture the nature of localized spin excitations in these strongly correlated systems. One potentially more accurate approach would be to calculate the energies associated with single spin excitations\cite{Tsuchiya-prl2006, Hsu-epsl, han-prb82, Hsu-prl} in these compounds. The results shown above, which demonstrate how the crystal and electronic structures can be properly reproduced in these materials, should form a solid basis for such calculations.

\subsection{Valence versus core treatment of \textit{R}CoO$_3$ 4\textit{f} electrons } \label{subsec:f-core-val}

Finally, we now investigate the effects of placing the rare-earth 4\textit{f} electrons into the core region of the pseudopotential for \textit{Pr}CoO$_3$, \textit{Gd}CoO$_3$ and \textit{Lu}CoO$_3$. These span the \textit{R}CoO$_3$ series, and we anticipate similar results for the intermediate compounds.

Due to the highly localized and strongly correlated nature of valence 4\textit{f} electrons, they need to be treated by methods beyond standard DFT, including DFT$+U$ or even hybrid functionals,\cite{hse,kresse_ce,hse_ce} making these rare-earth compounds computationally challenging. The DFT$+U$ method suffers from ambiguity in \textit{U} values and difficulty for computing $U_{sc}$ values for rare-earth 4\textit{f} electrons within the desired accuracy. On the other hand, hybrid functional calculations of  \textit{R}CoO$_3$'s are computationally demanding; it would be extremely difficult to obtain the results presented in the previous sections using hybrid functionals.  To circumvent these difficulties, we followed a similar method to that of Coh \textit{et. al.}\cite{sinisa} in which the 4\textit{f} electrons of rare-earth atoms are placed into the pseudopotential core so that the necessity for the DFT$+U$ method or other approaches addressing strong 4\textit{f}-correlations can be avoided. Obviously this approach is unable to describe phenomena involving magnetic ordering of rare-earth 4\textit{f} electrons at low temperatures, or other effects arising due to localized 4\textit{f} electrons. However, Coh \textit{et. al.}\cite{sinisa} showed that their approach is quite feasible within the limits of GGA overestimation\cite{overest} of structural properties of rare-earth scandates and yttrates in the Pbnm structure. In another study,\cite{bellaiche} the experimental Curie temperatures of bulk \textit{R$_2$}NiMnO$_6$ were shown to be well reproduced without the need to incorporate 4\textit{f} electrons in the valence states. Alternatively, one can ``push'' the 4\textit{f} electrons far away from Fermi level by applying large \textit{U} values, so that the structural and electronic properties of \textit{R}CoO$_3$'s related with Co 3\textit{d} states will not depend on 4\textit{f} electron details. This approach suffers from the difficulty of converging calculations, however. With 4 rare-earth ions in the unit cell, orbital and spin degrees of freedom in an incomplete \textit{f}-shell give rise to a huge number of inequivalent configurations with similar energies, and hence make calculations extremely difficult to converge using currently implemented mixing schemes in the Quantum ESPRESSO code. Due to such difficulties, we favor placing the 4\textit{f} electrons in the pseudopotential core, as long as their properties are not under scrutiny.

\begin{figure*}
\includegraphics[width=12cm]{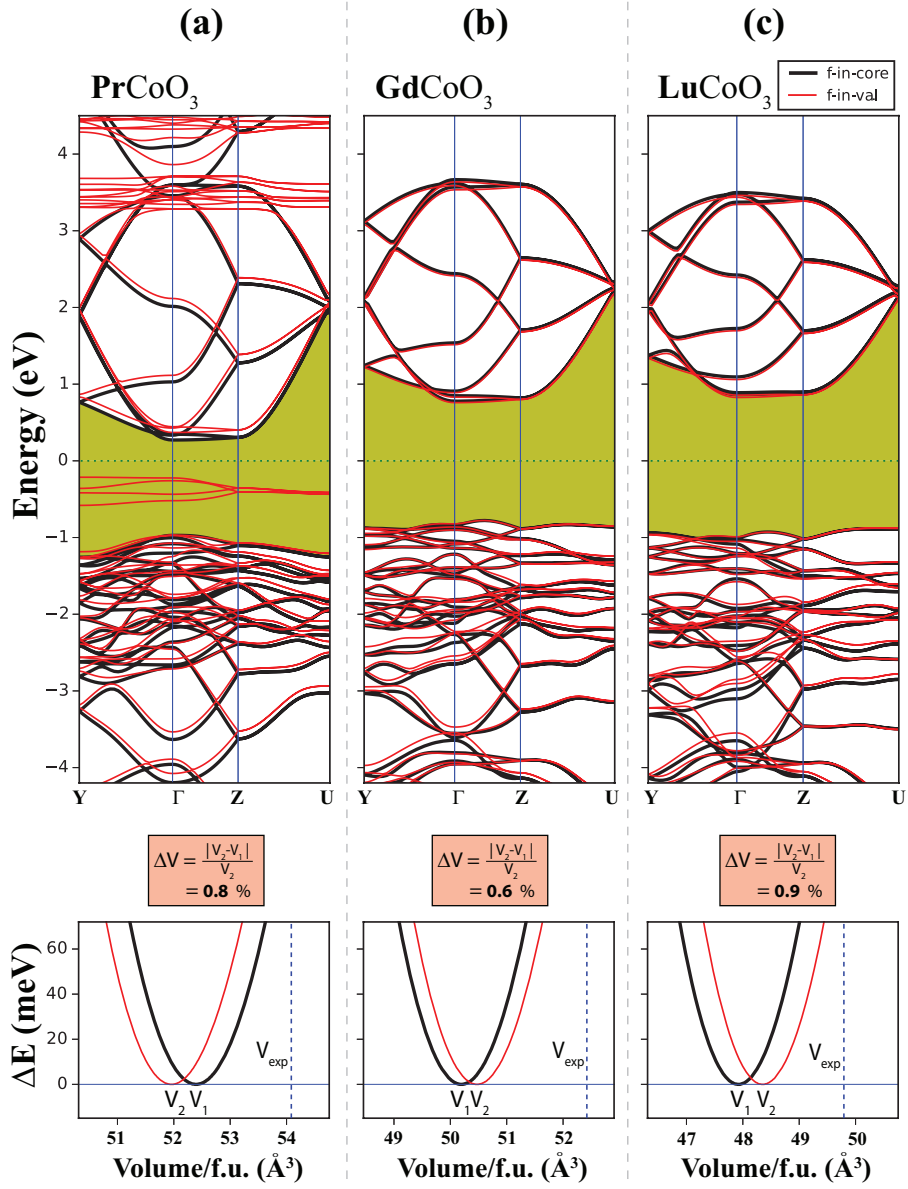}
\caption{(Color online) Comparison of band structure and energy versus volume curves of (a) \textit{Pr}CoO$_3$, (b) \textit{Gd}CoO$_3$ and (c)  \textit{Lu}CoO$_3$ in which 4\textit{f} electrons are kept as core (thick black lines) or valence (thin red lines) states. The Fermi level is set to 0 eV and the band gap is shaded. For clarity, bands along ``$Y - \Gamma - Z- U$`` path are shown. The changes are similar for other directions in Brillouin zone. The energy minimum is set to zero in the energy versus volume curve and experimental equilibrium volumes are indicated by a vertical dashed blue lines. The LDA+U$_{sc}$ method was used in the calculations.}
\label{fig:core-valence}
\end{figure*}

In Fig. ~\ref{fig:core-valence}, we compare band structures and the energy versus volume curves of \textit{Pr}CoO$_3$, \textit{Gd}CoO$_3$ and \textit{Lu}CoO$_3$ in which the 4\textit{f} electrons are kept as the core or valence states. These band structure calculations were performed at the experimental lattice constants using the LDA+U$_{sc}$ method. For the \textit{f}-in-valence case, 4.00, 4.60 and 5.50 eV Hubbard \textit{U} values for the rare-earth 4\textit{f}-electrons (as suggested in Ref.\onlinecite{mehmet-REN}) were used for  \textit{Pr}CoO$_3$, \textit{Gd}CoO$_3$ and \textit{Lu}CoO$_3$, respectively. 

Pr 3+ has electronic configuration \textit{[Xe]} 4\textit{f}$^2$. The occupied 4\textit{f} states in  \textit{Pr}CoO$_3$ form localized bands approximately in the middle of the gap, well separated from Co $e_{g}$-$t_{2g}$ and O \textit{p} states as shown in Fig. ~\ref{fig:core-valence}(a). Bringing Pr 4\textit{f} electrons into the valence states slightly increases (by $\sim$0.09 eV) the energy gap between Co $t_{2g}$-$e_{g}$ states, which was reported as $E_g$ in previous sections. The new \textit{f} to $e_{g}$ band gap is only 0.53 eV, significantly smaller than the $t_{2g}$-$e_{g}$ band gap  of 1.24 eV. The localized bands appearing around 4 eV in the conduction band originate from empty 4\textit{f} states. 

As seen in the lower panels in Fig. ~\ref{fig:core-valence}(a), the energy versus volume curve of \textit{Pr}CoO$_3$ slightly changes when the  4\textit{f} electrons are promoted to valence electrons. The change in equilibrium volume, $\Delta V = 0.8 \%$, is small compared to the LDA (GGA) underestimation (overestimation). For \textit{Gd}CoO$_3$ and  \textit{Lu}CoO$_3$,  the 4\textit{f} states are away from the Fermi level and the effects of promoting them to valence is quite small in the band structure; equilibrium volumes change by less than 1$\%$. These comparisons indicate that, except for the special case of \textit{Pr}CoO$_3$, electronic band structures calculated by placing the rare-earth 4\textit{f} states in the core are almost unchanged near the Fermi level. Equilibrium volumes are also relatively little affected compared to deviations between calculated and measured values.

The proximity of occupied Pr 4\textit{f} states to the Fermi level, as seen in Fig. ~\ref{fig:core-valence}(a), has been directly associated with the abrupt transport anomalies\cite{kanizek} and a temperature-dependent change of the Pr valence observed in compounds such as $Pr_{0.5}Ca_{0.5}$CoO$_3$.\cite{gonzales} On the other hand, its Nd counterpart ($Nd_{0.5}Ca_{0.5}$CoO$_3$) shows no such anomalies, though the ionic radius of Nd is only $\sim 1.6\%$ smaller than that of Pr. This difference in behavior is attributed to the shift of Nd 4\textit{f} states away from the Fermi level, deeper into the valence band, disabling changes in Nd 4\textit{f}  occupancies. \cite{kanizek, leightonNCO} As shown in Fig. ~\ref{fig:core-valence}  (b,c) this is also the case with \textit{Gd}CoO$_3$ and \textit{Lu}CoO$_3$ with 4\textit{f} states far from the Fermi level. Pr is thus unique in this respect, as has been previously acknowledged.

\section{Conclusions} \label{sec:conc}

In summary, we have performed first-principles calculations to investigate rare-earth cobaltite \textit{R}CoO$_3$ perovskites (\textit{R} = Pr -- Lu), including their structural and electronic properties. Several functionals have been tested. Among them, we found that LDA+$U_{sc}$ gives the most accurate results. For the structural properties of low-spin \textit{R}CoO$_3$, our calculations successfully capture the structural trends along the rare-earth series (\textit{R}=Pr -- Lu) observed in experiments, including the variation of equilibrium volume and Co-O-Co bond angles. Remarkably, the computed Co-O average bond length barely changes with the rare-earth elements. For electronic properties, we computed the band gaps, bandwidths, and crystal-field splittings of these cobaltites. The band gap increases along the series \textit{R}=Pr -- Lu, as in experiment. The evolution of the bandwidths and crystal field splittings, however, are contrary to simple expectations, demonstrating that prior simple rationalizations of the trend in spin crossover onset temperatures may be oversimplified.  We have also addressed the effects of placing the rare-earth 4\textit{f} electrons into the pseudopotential core. Given the accuracy and efficiency of our results and theoretical approach, we believe the Hubbard $U$ parameters presented in this work, and the DFT+$U_{sc}$ method, will enable additional predictive calculations for \textit{R}CoO$_3$ perovskites, eventually including the aforementioned trend in spin crossover onset temperature.

\section{Acknowledgments}
This work was supported primarily by the NSF MRSEC under award numbers DMR-0819885 and DMR-1420013. RMW was also supported by NSF/EAR 1319361. CL's contribution was supported by the DOE through DE-FG02-06ER46275. We acknowledge fruitful discussions with Dr. Koichiro Umemoto and Dr. Han Hsu. Part of the computational resources have been provided through the Extreme Science and Engineering Discovery Environment (XSEDE), supported by National Science Foundation grant number ACI-1053575.

\section{Appendix A: Access to rare-earth potentials}
LDA and GGA versions of rare-earth potentials used in this study are publicly available at http://www.vlab.msi.umn.edu/resources/repaw/index.shtml.

\pagebreak

\pagebreak

\pagebreak

\pagebreak

\end{document}